%% file: main.tex
\documentclass{vldb}
\pdfoutput=1

\usepackage{times}
\usepackage{fullpage}
\usepackage{graphicx}
\usepackage{balance} 
\usepackage{color}
\usepackage[utf8]{inputenc}
\usepackage{listings} 
\usepackage{comment}
\usepackage{mdwlist}
\usepackage{tabularx} 
\usepackage{tikz}
\usepackage{url}
\usepackage{color}
\usepackage{xspace}
\usepackage{appendix}
\usepackage{cleveref}
\usepackage[binary-units]{siunitx}
\usepackage{paralist}
\usepackage{todonotes}
\usepackage{enumitem}
\usepackage{pgfplots}
\usepackage{tikz}
\usepackage{subfig}
\usepackage{hyperref}

\usepackage[skip=5pt]{caption}

\newcommand{\system}{DBPal\xspace}
\newcommand{\naive}{na\"{\i}ve\xspace}
\newcommand{\Naive}{Na\"{\i}ve\xspace}

\begin{document}

\title{\system{}: An End-to-end Neural\\ Natural Language Interface for Databases}

\numberofauthors{1}

\author{
\alignauthor 
Prasetya Utama\textsuperscript{1},
Nathaniel Weir\textsuperscript{1},
Fuat Bas{\i}k\textsuperscript{3}, 
Carsten Binnig\textsuperscript{1,2},
Ugur Cetintemel\textsuperscript{1},
Benjamin H\"attasch\textsuperscript{2},
Amir Ilkhechi\textsuperscript{1},
Shekar Ramaswamy\textsuperscript{1},
Arif Usta\textsuperscript{3}
\and 
\begin{tabular}{ccc}\\
\textsuperscript{1} Brown University, USA &
\textsuperscript{2} TU Darmstadt, Germany &
\textsuperscript{3} Bilkent University, Turkey
\end{tabular}
}

\maketitle

\begin{abstract} 
The ability to extract insights from new data sets is critical for decision making.
Visual interactive tools play an important role in data exploration since they provide non-technical users with an effective way to visually compose queries and comprehend the results.
Natural language has recently gained traction as an alternative query interface to databases with the potential to enable non-expert users to formulate complex questions and information needs efficiently and effectively.
However, understanding natural language questions and translating them accurately to SQL is a challenging task, and thus Natural Language Interfaces for Databases (NLIDBs) have not yet made their way into practical tools and commercial products.

In this paper, we present \system{}, a novel data exploration tool with a natural language interface. \system{} leverages recent advances in deep models to make query understanding more robust in the following ways:
First, \system{} uses a deep model to translate natural language statements to SQL, making the translation process more robust to paraphrasing and other linguistic variations.
Second, to support the users in phrasing questions without knowing the database schema and the query features, \system{} provides a learned auto-completion model that suggests partial query extensions to users during query formulation and thus helps to write complex queries.
\end{abstract}

\input{01_intro.tex}
\input{02_overview.tex}
\input{03_training.tex}
\input{04_runtime.tex}
\input{05_complex.tex}
\input{06_evaluation.tex}
\input{07_userstudy.tex}
\input{08_related.tex}
\input{09_conclusion.tex}

\begin{scriptsize}
\bibliographystyle{abbrv}
\bibliography{bib}  
\end{scriptsize}
\end{document}

%% file: 01_intro.tex
\section{Introduction}

\label{sec:intro}
\textbf{Motivation:} Structured query language (SQL), despite its expressiveness, may hinder users with little or no relational database knowledge from exploring and making use of the data stored in an RDBMS. In order to effectively leverage their data sets, users are required to have prior knowledge about the schema information of their database, such as table names, columns and relations, as well as a working understanding of the syntax and semantics of SQL. These requirements set ``a high bar for entry" for democratized data exploration and thus have triggered new research efforts to develop alternative interfaces that allow non-technical users to explore and interact with their data conveniently.
While visual data exploration tools have recently gained significant attention, Natural Language Interfaces to Databases (NLIDBs) appear as highly promising alternatives because they enable users to pose complex ad-hoc questions in a concise and convenient manner.

\begin{figure}
\centering
\vspace{4ex}
\includegraphics[width=0.5\textwidth]{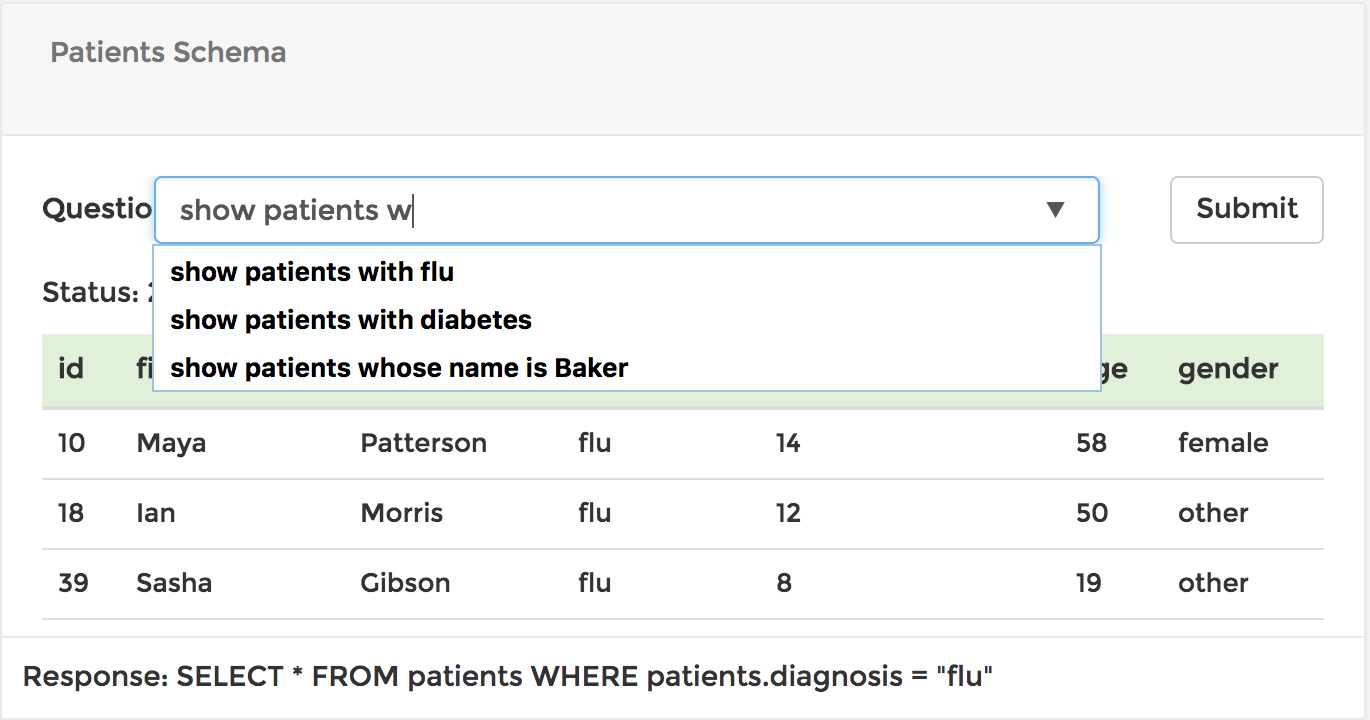}
\vspace{2.5ex}
\caption{An Example Session in \system{}}
\vspace{-3.5ex}
\label{fig:screenshot}
\end{figure}

For example, imagine that a medical doctor starts her new job at a hospital and wants to find out about the age distribution of patients with the longest stays in the hospital.
This question typically requires the doctor -- when using a standard database interface directly -- to write a complex nested SQL query.
Even with a visual exploration tool such as Tableau \cite{tableau_sigmod15} or Vizdom \cite{vizdom_vldb15}, a query like this is far from being trivial since it requires the user to execute multiple query steps and interactions.
Alternatively, with an exploration tool supported by a natural language interface, the query would be as simple as stating ``What is the age distribution of patients who stayed longest in the hospital?''\\

\textbf{Contribution:} In this paper, we introduce \system{}, a relational database exploration tool that provides a robust and easy-to-use natural language (NL) interface with the purpose of improving the transparency of the underlying database \linebreak schema and enhancing the expressiveness and flexibility of human-data-interaction.
Different from existing approaches, \system{} leverages deep neural network models as the core of its natural language interface system.
In the following, we outline the two key features of \system{} that are based on deep neural network models.

\emph{Robust Query Translation:} We propose a novel query translation framework based on a sequence-to-sequence recurrent neural network model that has recently became a state-of-the-art for machine translation task. 
Our notion of model robustness is defined as the effectiveness of the translation model to map linguistically varying utterances to finite predefined relational database operations. Take, for example, the SQL expression \emph{SELECT * FROM patients} \emph{WHE\-RE diagnosis='flu'}. There are numerous corresponding natural language utterances for this query, such as \emph{"show all patients with diagnosis of flu"} or simply \emph{"get flu patients"}. We aim to build a translation system that is invariant towards these linguistic alterations, no matter how complex or convoluted. 

A key challenge hereby is to curate a comprehensive training set for the model. While existing approaches for machine translation require a manually annotated training set, 
we implement a novel synthetic generation approach that uses only the database schema with minimal annotation as input and generates a large collection of pairs of natural language queries and their corresponding SQL statements.
In order to maximize our coverage towards natural language variation, we augment the training set using a series of techniques, among which is an automatic paraphrasing process using an existing 
paraphrasing database called PPDB \cite{DBLP:conf/acl/PavlickC16a} extracted from large text corpora. 

\emph{Interactive Auto-Completion:} We provide real-time auto-completion and query suggestions to help users who may be unfamiliar with the database schema or the supported query features.
This helps to improve translation accuracy by leading the user towards less ambiguous queries. Consider a scenario in which a user is exploring a US geographical database and starting to type \emph{"show me the names~"} -- at this point, the system suggests possible completions such as \emph{of states}, \emph{of rivers}, or \emph{of cities} to make the user aware of the different options she has, given the specific database context. At the core of the auto-completion feature is a language model based on the same sequence-to-sequence architecture and trained on the same synthetic training set as the query translator.

As a result, \system{} is the \textbf{first system that allows users to build a robust NL-interface for a new database schema with only minimal manual annotation overhead}. 
A screenshot of our prototype of \system{}, which implements the aforementioned features, is shown in Figure \ref{fig:screenshot}.
We encourage readers to watch the
video\footnote{\url{https://vimeo.com/user78987383/dbpal}}, which shows a recording of a representative user session with our system.
In this paper, our main focus will be on the first component, the NL-to-SQL translator. We will also briefly describe the interactive auto-completion process for completeness.\\

\textbf{Outline:} The remainder of this paper is organized as follows: 
in Section \ref{sec:overview}, we introduce the overall system architecture of \system{}. 
Afterwards, in Section \ref{sec:training} we describe the details of the training phase of \system{} for the NL-to-SQL translation.
We then show how the learned model for NL-to-SQL translation is applied at runtime in Section \ref{sec:runtime}.
Furthermore, we discuss the handling of more complex queries like joins and nested queries in Section \ref{sec:complex}. 
In order to show the efficiency of \system{} as well as its robustness, we present our evaluation results using benchmarks in Section \ref{sec:evaluation} as well as the results of a user study in Section \ref{sec:study}. 
Finally, we discuss related prior work in Section \ref{sec:related} and 
then conclude by discussing planned future extensions in Section \ref{sec:concl}.

%% file: 02_overview.tex
\section{Overview}
\label{sec:overview}

\subsection{System Architecture}

\begin{figure}
\centering
\includegraphics[width=0.45\textwidth]{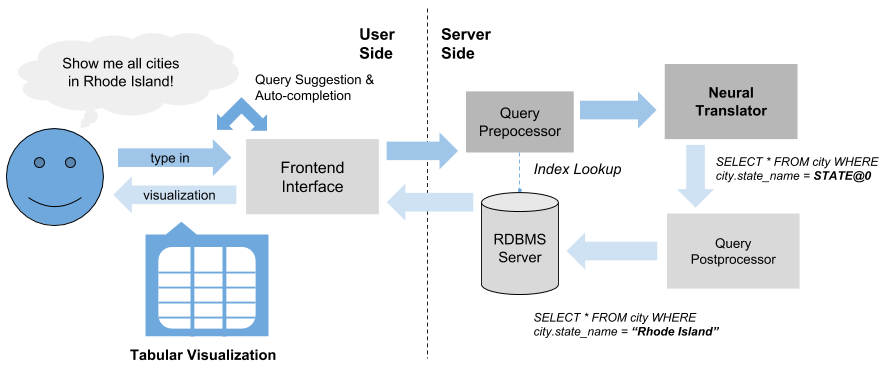}

\vspace{2ex}
\caption{System Overview of \system{} 
\vspace{-3.5ex}
}
\label{fig:system}
\end{figure}

As shown in Figure \ref{fig:system}, the \system{} system consists of two main components: 
a neural query translator and an interactive auto-completion feature to help users to phrase questions.

\paragraph*{Neural Query Translation}
The main task of the query translation is to generate SQL queries from a natural language (NL) query.
The main novelty of the translation process is how we pose the task as a language translation problem, which is then modeled using the state-of-the-art Recurrent Neural Network (RNN) architecture \cite{Sutskever:2014:SSL:2969033.2969173}.
A major challenge when using such a deep model is the need for a large and comprehensive training set consisting of Natural-Language-to-SQL (NL-SQL) pairs. 
Existing work \cite{DBLP:conf/acl/IyerKCKZ17} therefore requires huge efforts because a manually curated training corpus is needed to train a NL-SQL deep model for every new database schema that needs to be supported.

Our approach is very different since we do not need such manual effort.
The main idea is that we can ``generate'' a training set for a given database schema fully automatically as shown in Figure \ref{fig:translator} (left-hand-side).
We train an RNN-model with this generated data and use it at runtime to translate NL queries to SQL as shown in Figure \ref{fig:translator} (right-hand-side).
For the query translation at runtime, \system{} applies some additional pre- and post-pro\-cessing.
In these steps, we handle parameters, correct SQL syntax errors, and expand complex queries such as SQL joins.

\paragraph*{Interactive Auto-Completion}
To improve the search performance of users, \system{} also provides an auto-completion mechanism for NL queries. 
This not only enhances the search behavior, but also contributes towards increased translation accuracy by leading the user to submit less ambiguous queries. 
Considering the unawareness of users regarding the database schema, auto-completion becomes even more crucial.
The core of the auto-completion component is comprised of the same type of sequence-to-sequence learning model trained on the same training set as the query translation model. 
At runtime, the model is used together with a modified search algorithm. 
Given an inputted NL query, the auto-completion model performs a breadth-first search in the candidate space of possible SQL translations and suggests to the user the most probable ways to complete potential queries.
In the remainder of this paper, we will focus the discussion on the first component only, which is the NL-to-SQL translation.

\subsection{Training Data Generation}

As mentioned above, the most important aspect of \system{} is that given a database schema, we automatically generate the data set to train the translation model.
In the following, we give a brief overview of our data generation framework.

\begin{figure}
\centering
\includegraphics[width=0.45\textwidth]{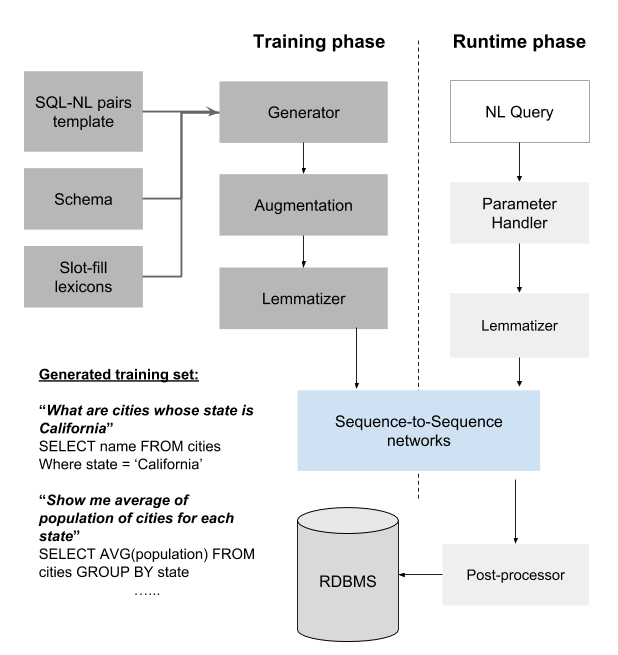}
\vspace{2ex}
\caption{\system{}'s Neural Query Translator 
\vspace{-3.5ex}}
\label{fig:translator}
\end{figure}

In the first data generation step (called \emph{Generator} in Figure \ref{fig:translator}), we use the database schema along with a set of base templates that describe NL-SQL pairs and so called `slot-filling' dictionaries to generate a training set of 1-2 million pairs. 
While the schema information is used to instantiate table and attribute names in the templates, the slot-fill dictionaries are used to instantiate different variations of NL words and phrases.

In the second step, we automatically augment the initial set of NL-SQL pairs by leveraging existing language models to automatically vary the NL-part of each pair using different linguistic variations (called \emph{Augmentation} in Figure \ref{fig:translator}).
The goal of the augmentation phase is to cover a wide spectrum of linguistic variations of how users might query the given database schema.
This augmentation is the key to make the translation model robust and allows \system{} to provide better query understanding capabilities than other existing NL-interfaces for databases. 

Finally, in the last step of the data generation procedure the resulting NL-SQL pairs are lemmatized to normalize the representation of individual words.
The same lemmatization step is also applied at runtime during the pre-processing phase.
In our experiments, we saw that lemmatization of NL queries at training and runtime increased the accuracy of the translation.

%% file: 03_training.tex
\section{Training Phase}
\label{sec:training}

The translation of NL to SQL can be thought of as a function $T$ that maps a given sequence of words $s$ (which describes the input NL query) into another sequence $s^\prime$  --- the output SQL query. 
In this paper, we therefore use a deep sequence-to-sequence (seq2seq) model to implement this complex mapping of $T$. 
However, using a deep seq2seq model for the translation from NL to SQL necessitates a comprehensive training dataset comprised of NL-SQL pairs.
Furthermore, the NL-SQL pairs used to learn the translation function $T$ depend heavily on the database schema.

Since one of the main promises of \system{} is that we require minimal manual overhead during training data curation to create an NL interface for any new database schema, we provide a novel way to synthetically generate the training data set.
The complete process of the training phase including the data generation is shown in Figure \ref{fig:translator} (left-hand-side).

In the following, we first discuss the details of our data generation approach (which is part of our training phase that consists of two steps: data instantiation and data augmentation).
Afterwards, we discuss further optimization of the data generation process to increase the model quality based on automatic parameter tuning.
Finally, in the last part of this section, we elaborate on the concrete architecture we use for our translation model including the hyperparameters used to train the model.

\subsection{Data Instantiation}

The observation is that SQL, as opposed to natural language, has a much more limited expressiveness and therefore multiple utterances with the same semantic meaning might be mapped to the same SQL query. 
For this section, we assume that all queries in the training set are using only one table to simplify the presentation.
More complex queries including joins and nested queries are discussed later in Section \ref{sec:complex}.

The main idea of the data instantiation is that the space of possible SQL queries ($S^\prime$) a user might phrase against a given database schema is defined using query templates.
These query templates are used to instantiate different possible SQL queries. 
An example SQL template is as follows:\\

\makebox[0.48\textwidth][c]{
\centering
\begin{minipage}[c]{0.48\textwidth}
\noindent \textbf{SQL Template:} \\
\textit{Select \textsf{\{Attribute\}(s)} From \textsf{\{Table\}} Where \textsf{\{Filter\}}}\\
\end{minipage}}

The SQL templates we use in \system{} cover a variety of different types of queries ranging from select-from-where queries to aggregate-grouping queries.
For each SQL template, we also provide one or more natural language (NL) templates as counterparts that are used for the instantiation of NL-SQL pairs such as:\\

\makebox[0.48\textwidth][c]{
\centering
\begin{minipage}[c]{0.48\textwidth}
\noindent \textbf{NL Template:} \\
\textit{\{SelectPhrase\} the \textsf{\{Attribute\}(s)} \{FromPhrase\} \textsf{\{Table\}} 
\{WherePhrase\} \textsf{\{Filter\}}}\\
\end{minipage}}

In order to reflect the larger expressiveness of NL versus SQL, our NL templates define slots that allow us to instantiate different NL variations (e.g., $SelectPhrase$, $FromPhrase$, $WherePhrase$) in addition to the slots for database items (e.g., $Table$, $Attributes$, $Filter$) present in the NL and SQL templates.

To initially create the training set, the generator repeatedly instantiates each pair of SQL and corresponding NL templates by filling in their slots. 
Table and attribute slots are filled in using the schema information of the database, while a diverse array of natural language slots is filled using a manually-crafted dictionary of synonymous words and phrases. 
For example, the $SelectPhrase$ can be instantiated using \emph{What is} or \emph{Show me}, among others.
Thus, an instantiated SQL-NL pair might look like this:\\

\makebox[0.48\textwidth][c]{
\centering  
\begin{minipage}[c]{0.48\textwidth}
\noindent \textbf{Instantiated SQL Query:} \\
\emph{SELECT name FROM patient WHERE age=@AGE}
\vspace{1.5ex}\\
\textbf{Instantiated NL Query:} \\
\emph{Show me the name of all patients with age @AGE}\\
\end{minipage}}

It is important to note that we do not use actual constants in the filter predicates. 
Instead, we use placeholders that represent an arbitrary constant for a given table attribute.
This makes the model (that is trained on the generated training data) independent of concrete values used in the database; thus retraining is not required after inserts or updates.

It is also of importance to balance the training data when instantiating the slots with possible values.
If we naively replace the slots of a NL-SQL template pair with all possible combinations of slot instances (e.g., all attribute combinations of the schema), then instances resulting from certain templates with more slots would dominate over instances from templates with fewer slots (given that the addition of a new slot exponentially increases the number of instances that can be generated) and thus add bias to the translation model for certain query classes.
An imbalance of the instances could result in a biased training dataset where the trained model would prefer certain translations over other ones only due certain variations appearing more often. 
We therefore provide a parameter $size_{slotfills}$ to set a maximum number of instances created for each NL-SQL template pair.
If the possible combinations of slot instantiations exceed $size_{slotfills}$, we randomly sample from the possible combinations. 

Moreover, for one SQL template we typically provide multiple corresponding NL templates that follow particular paraphrasing techniques as discussed in \cite{DBLP:journals/pdln/VilaMR11}, covering categories such as syntactical, lexical or morphological paraphrasing. An example of such a manually paraphrased NL template (representing a syntactical variation of the last example) and its instantiation is shown next:\\

\makebox[0.48\textwidth][c]{
\centering
\begin{minipage}[c]{0.48\textwidth}
\noindent \textbf{Paraphrased NL Template:} \\
\emph{For \textsf{\{Table\}(s)} with \textsf{\{Filter\}}, what is their \textsf{\{Attribute\}(s)}?}
\vspace{1.5ex}\\
\textbf{Instantiated NL Query (Paraphrased):} \\
\emph{For patients with age @AGE, what is their name?}\\
\end{minipage}}

All the SQL and NL query templates for the data instantiation phase in \system{} are crafted manually -- however, these templates are all database schema independent and can be applied to instantiate NL-SQL pairs for any possible schema without additional manual effort.
In the current version of \system{}, we provide a set of approximately $90$ different templates to instantiate pairs of SQL-NL queries in the first step.
After the instantiation step, all the generated SQL-NL query pairs are handed over to the automatic augmentation step that additionally applies automatic paraphrasing techniques among some other steps as we will discuss next.

\subsection{Data Augmentation}

In order to make the query translation model robust against linguistic variations of how a user might phrase an input NL query, we apply the following augmentation steps for each instantiated NL-SQL pair:

\paragraph*{Automatic Paraphrasing}
First, we augment the training set by duplicating NL-SQL pairs, but randomly selecting words and sub-clauses of the NL query and paraphrasing them using The Paraphrase Database (PPDB) \cite{DBLP:conf/acl/PavlickC16a} as a lexical resource. An example for this is:\\

\makebox[0.48\textwidth][c]{
\centering
\begin{minipage}[c]{0.48\textwidth}
\noindent \textbf{Input NL Query:} \\
\emph{\underline{Show} the name of all patients with age @AGE}
\vspace{1.5ex}\\
\noindent \textbf{PPDB Output:} \\
demonstrate, showcase, display, indicate, lay
\vspace{1.5ex}\\
\noindent \textbf{Paraphrased NL Query:} \\
\emph{display the names of all patients with age @AGE}\\
\end{minipage}}

PPDB is an automatically extracted database containing millions of paraphrases in 16 different languages. 
In its English corpus which we currently use in \system{}, PPDB provides over 220 million paraphrase pairs, consisting of 73 million phrasal and 8 million lexical paraphrases, as well as 140 million paraphrase patterns, which capture many meaning-preserving syntactic transformations. 
The paraphrases are extracted from bilingual parallel corpora totalling over 100 million sentence pairs and over 2 billion English words. 

In the automatic paraphrasing step of \system{}, we use PPDB as an index that maps words/sub-clauses to paraphrases and replace words/sub-clauses of the input NL query with available paraphrases.
For any given input phrase to PPDB, there are often dozens or hundreds of possible paraphrases available.
For example, searching in PPDB for a paraphrase of the word \emph{enumerate} which can be used as a part of the question \emph{"enumerate the names of patients with age 80"}, we get suggestions such as \emph{list} or \emph{identify}.

An important question is how aggressively we apply automatic paraphrasing. 
We therefore provide two parameters to tune the automatic paraphrasing in \system{}. The first parameter $size_{para}$ defines the maximal size of the sub-clauses (in number of words) that should be replaced in a given NL query. A  second parameter $num_{para}$ defines the maximal number of paraphrases that are generated as linguistic variations for each sub-clause.
For example, setting $size_{para}=2$ will replace sub-clauses of size $1$ and $2$ (i.e., uni-grams and bi-grams) in the input NL query with paraphrases found in PPDB.  
Furthermore, setting $num_{para}=3$, each of the uni- and bi-grams will be replaced by at most $3$ paraphrases. 

Setting these two parameters in an optimal manner is, however, not trivial: 
if we set both parameters to high values, we can heavily expand our initial training set of NL-SQL query pairs using many different linguistic variations, which hopefully will increase the overall robustness of \system{}.
However, as a downside, at the same time we might also introduce noise into the training data set since PPDB also includes paraphrases with low quality. 
In order to find parameter values that increase the overall model accuracy for a given input database schema, we provide an optimization procedure that we discuss in Section \ref{sec:training:optimization}. It automatically finds a parameterization that balances among others the trade-off between these two dimensions: size of the augmented training data and noise in the training data. 

\paragraph*{Missing Information}
Second, another challenge of input NL queries is missing or implicit information.
For example, a user might ask for \emph{patients with flu} instead of \emph{patients diagnosed with flu} and thus the information about the referenced attribute might be missing in a user query.

Therefore, to make the translation more robust against missing information, we duplicate individual NL-SQL pairs and select individual words/sub-clauses that are removed from the NL query. 
Similar to the paraphrasing process, an interesting question is which words/sub-clauses should be removed and how aggressively the removal should be applied.

Currently, we follow a similar protocol as the paraphrasing process by randomly selecting words in the NL query and removing them in a duplicate. 
Again, aggressively removing words from query copies increases the training data size since more variations are generated.
On the other hand, however, we again might introduce noise; when removing too many words, similar-looking NL queries might then translate into incorrect SQL queries.
In order to tune how aggressively we apply removing, we additionally provide a  parameter named $num_{missing}$ that defines the maximum number of query duplicates with removed words for a given input NL query. We also include a parameter $randDrop_p$ that defines how often the generator will choose to remove words from a particular NL query at all.
Analogously to automatic paraphrasing, we set these two parameters for a given input database schema automatically as described in Section \ref{sec:training:optimization}.  

\paragraph*{Other Augmentations}
For the automatic data augmentation, we apply some additional techniques to increase the linguistic variations.
One example is the usage of available linguistic dictionaries for comparatives and superlatives.
That way, we can replace for example the general phrase \emph{greater than} in an input NL query by \emph{older than} if the domain of the schema attribute is set to \emph{age}.

In the future, we plan on extending our augmentation techniques; e.g., one avenue is to enhance our automatic paraphrasing using other language sources and not only PPDB. 
We also plan to investigate the idea of using an off-the-shelf part-of-speech tagger to annotate each word in a given NL query.
These annotations can be used in different forms: for example, we could use them in the automatic paraphrasing to identify better paraphrases or to infer a request for a nested query.
We could also use them to apply the word removal only for certain classes of words (e.g., nouns).

\subsection{Optimization Procedure}
\label{sec:training:optimization}
 
One important challenge of the automatic data generation steps is to instantiate the training data such that the translation model will provide a high accuracy. 
For example, the data augmentation steps discussed above require several parameters for each step that define how aggressively paraphrasing and removing information is applied to an input NL query.
Furthermore, the template-based training data instantiating step also has parameters that can be tuned to balance the number of NL-SQL pairs that are instantiated for each template. We therefore attempt to automatically optimize the  configuration of generator parameters given a particular database schema. 

The intuition behind this strategy derives from observations we made about the seq2seq model's behavior -- in particular that it is very susceptible to fitting to over-represented words or functions. For example, if we overpopulate the training set with the SQL \emph{count} operative (the natural language parallel will usually include words like `how many' then), the model will be prone to output a \emph{count} operative simply because it sees particular NL words that most commonly appeared with \emph{count}s during training. Queries like `\emph{how} large is the area of Alaska' might be mapped to a \emph{count} instead of  \emph{sum} simply for this reason. It is the goal of the optimization process to negate as many of these  translation mishaps as possible simply by making adjustments to the query representations.

Table \ref{table:parameters} lists all parameters that are available in \system{} to tune the data generation process and explains their meaning:

\begin{table}[]
\small
\begin{center}
\begin{tabular}{| p{2cm} | p{5cm} |}\hline
\textbf{Parameter} & \textbf{Explanation} \\ \hline
\multicolumn{2}{|c|}{\textbf{Data Instantiation}} \\ \hline
$size_{slotfills}$ & Maximal number of instances created for a NL-SQL template pair using slot-filling dictionaries.\\ \hline
$size_{tables}$ & Maximum number of tables supported in join queries.\\ \hline $groupBy_{p}$ & Probabilities of generating a \emph{GROUP BY} version of a generated query pair.\\ \hline
$join_{boost}$, $agg_{boost}$, $nest_{boost}$ & Control the balance of various types of SQL statements relative to each other and the number of templates used.\\ \hline
\multicolumn{2}{|c|}{\textbf{Data Augmentation}} \\ \hline
$size_{para}$ & Maximal size of sub-clauses that are automatically replaced by a paraphrase.\\ \hline
$num_{para}$ & Maximal number of paraphrases that are used to vary a sub-clause. \\ \hline
$num_{missing}$ & Maximal number of words that are removed for a given input NL query.\\ \hline
$randDrop_p$ & Probability of randomly dropping words from a generated query.\\ \hline
\end{tabular}
\end{center}
\vspace{-2ex}
\caption{Tuning Parameters of the Data Generation Procedure}
\vspace{-2ex}
\label{table:parameters}
\end{table}

As mentioned before, the parameters listed in the table above define the main characteristics of the training data instantiation and augmentation step and thus have an effect on the accuracy of the seq2seq-translation model of \system{} trained on this data set.
In order to find the optimal parameter setting for the data generation process automatically, we model the data generation procedure as the following function:

$$Acc = Generate(D, T, \phi{})$$

The input of this function are the database $D$ that describes the schema and contains some sample data, a test workload $T$ of input NL queries and expected output SQL queries, as well as a possible instantiation of all the tuning parameters $\phi{}$ listed in the table before.  
The output of the generation procedure $Acc$ is the accuracy of our model that is trained on the generated data set using $D$ as well as $\phi{}$ and then evaluated using the test workload $T$. 
It is important to note that we can either use a test workload $T$ that is created automatically by using a random sample of the generated training data (i.e., we split the test set from the training data set) or by providing a small representative set of NL and SQL queries that are curated manually.

In order to optimize the parameters $\phi$ for the generation procedure, we use Bayesian Optimization as a sequential design strategy to optimize the parameters $\phi$ of the black-box function $Generate$ since it does not require derivatives.
The goal of the Bayesian optimization is to find a parameter set $\phi$ which maximizes the accuracy $Acc$.
To that end $\phi$ could also be seen as some hyperparameters of the generation function $Generate$.
In the experimental evaluation, we show that by using our optimization procedure, we can find parameters for the data generation process to produce training data that can provide a high accuracy for the trained model.

\subsection{Model Architecture}

Our translation model follows a similar architecture of a sequence-to-sequence (seq2seq) model in \cite{Sutskever:2014:SSL:2969033.2969173} for machine translation tasks. 
The model itself consists of two recurrent neural networks, namely, the encoder and decoder. 
We use a bidirectional encoder-decoder architecture as proposed by \cite{DBLP:journals/corr/BahdanauCB14}. 

Both the encoder and decoder comprise of two layers of gated recurrent units (GRUs). 
The dimension of the hidden state vectors as well as the the word embeddings is currently set to be $500$ and $300$ respectively, which we also found by automatic hyperparameter tuning. 
During training, we additionally apply a dropout on the embedding layer to avoid overfitting to our training corpus, which only consists of a small vocabulary  (i.e., SQL keywords as well as schema information of the given database).

%% file: 04_runtime.tex
\section{Runtime Phase} 
\label{sec:runtime}

In this section we describe the query translation pipeline. The complete process of the runtime phase is shown in Figure \ref{fig:translator} (right-hand-side).
From the given input natural language query to the output SQL query three major processing phases take part: pre-processing, query-translation and post-processing.
The output SQL query is then executed against the database and the result is returned to the user interface of \system{}.

\subsection{Pre-processing and Query Translation}

The input of the pre-processing step is a NL query formulated by the user such as the following example:\\

\makebox[0.48\textwidth][c]{
\centering
\begin{minipage}[c]{0.48\textwidth}
\noindent 
\textbf{User NL Query (with constants):} \\
\emph{Show me the name of all patients with age 80}\\
\end{minipage}}

In the first step (the pre-processing phase) parameter values (i.e., constants) are replaced with special placeholders.
This is due to the fact that the training dataset also does not use concrete parameter values in order to be able to translate queries independently from the database content as follows:\\

\makebox[0.48\textwidth][c]{
\centering
\begin{minipage}[c]{0.48\textwidth}
\noindent \textbf{Input NL Query (without constants):} \\
\emph{Show me the name of all patients with age @AGE}
\vspace{1.5ex}\\
\textbf{Output SQL Query (without constants):} \\
\emph{SELECT name FROM patient WHERE age=@AGE}\\
\end{minipage}}

Replacing the constants in the input NL query with their placeholders is a non-trivial task. 
In the basic version of \system{}, we therefore build an index on each attribute of the schema that maps constants to possible attribute names.
However, the basic version has several downsides: 
First, the mapping is not unique since the same constant might map to different columns.
Second, the user might use linguistic variations of constants in the database (e.g., ``NYC'' instead of ``New York City'') or the user might query for constants that are not present in the database which is often the case when queries use negations (e.g. NOT IN ``NYC'').

In order to better support the mapping from a constant to a column name, we additionally use a model-based approach in \system{} if the basic approach is not able to perform the mapping.
In order to do that we leverage word embeddings \cite{DBLP:journals/corr/abs-1301-3781} to map a given constant to a column name. 
For each column, the average vector representation of the values inside the column is calculated. 
Then, the constant to be labeled is compared against all of the average vector representations of all columns of the database schema. 
The closest one in terms of vector space is chosen to be mapping given it is bigger than a predefined threshold.  

Finally, as a last step of pre-processing we lemmatize the input NL query to normalize individual words and thus increase the similarity of the training data (which we also lemmatized) and the input NL query a user provides at runtime. 
After all pre-processing steps are applied, the trained model is used to map the lemmatized NL query where constants are replaced with placeholders into an output SQL query as shown before already.

\subsection{Post-Processing}

After pre-processing and translation, a post-processing phase is applied. 
The main two actions of post-processing are: 
first, the placeholders are substituted by the constants in the output SQL query.
Second, we use SQL syntax knowledge to repair potential translation errors of the model.

The first step is quite simple, since it implements the reverse step of the pre-processing phase. For example, the placeholder in the output SQL query shown before should be replaced by the according constant that was present in the input of the user:\\

\makebox[0.48\textwidth][c]{
\centering
\begin{minipage}[c]{0.48\textwidth}
\noindent 
\textbf{Output SQL Query (with constants):} \\
\emph{SELECT name FROM patient WHERE age=80}\\
\end{minipage}}

Hence we need to replace the placeholder in the SQL output of the model with the constant used in the input NL query  (e.g., @AGE is replaced by the constant $80$ in the example above). 
For string constants, this is not trivial since the user might have provided a string constant in the input NL query which is only similar to the one used in the database (e.g., the user provides ``New York City'' instead of ``NYC'').
In the current version of \system{}, we use a similarity function to replace constants with their most similar value that is used in the database.
We therefore search the possible column values and compute a string similarity metric with the string constant provided by the user.
In \system{}, we currently use the Jaccard-distance but the function can be replaced with another similarity metric.
In case the similarity of all values with the user given string is too low (which could mean that the value does not exist in the database), we use the constant as given by the user and do not replace it. 

In the second step of the post-processing phase \system{} uses knowledge about the SQL syntax to repair potential translation errors that might result from applying our seq2seq model.
One typical example for this is that the attributes used in the output SQL query and the table names do not match (e.g., the query asks for patient names but the table patient is not used in the FROM-clause).
In this case, the post-processing step adds missing tables to the FROM-clause.
The most likely join path is selected by the schema information using the shortest join path between the table already used in the FROM-clause and the missing table.
This is similar to the general join handling, which we discuss in detail in the next section.

%% file: 05_complex.tex
\section{Complex Queries}
\label{sec:complex}

In the previous sections, we have presented how the training and runtime phase of \system{} works for queries with single tables.
In this section, we discuss the extensions of \system{} to handle join and nested queries, too.

\subsection{Join Queries}

In order to handle NL input queries that require a SQL join in the output, we extend \system{} phase as follows.

For the training phase, we extend the template-based instantiation such that the attribute slots of a query can be filled with attribute names from different tables.
Attribute slots can be present in different parts of a query; e.g., the SELECT or the FILTER clause.  
The maximum number of distinct tables that are used during slot-filling can be defined using a parameter called $size_{tables}$, which is a tuning parameter of the data generation process as discussed before.
Furthermore, we also change the instantiation of table names in the generated SQL query.
Instead of enumerating all required tables in the FROM-clause, we add a special placeholder \textit{@JOIN}.
An example for an instantiated NL- and SQL-query that use a join could thus look as follows:\\

\makebox[0.48\textwidth][c]{
\centering
\begin{minipage}[c]{0.48\textwidth}
\noindent \textbf{SQL Query (Training Set):} \\
\emph{SELECT AVG(patient.age) FROM @JOIN WHERE \\doctor.name=@DOCTOR.NAME}
\vspace{1.5ex}\\
\textbf{NL Query (Training Set):} \\
\emph{What is the average age of patients treated \\by doctor @DOCTOR.NAME}\\
\end{minipage}}

At runtime, our translation model then outputs a SQL query with a \textit{@JOIN} placeholder when it sees an input NL query with attributes from multiple tables; i.e., it outputs a SQL query without concrete table names in the FROM-clause.
The \textit{@JOIN} placeholder is then replaced in the post-processing step with the actual table names and the join path that contains all tables required by the query.
We saw that this reduces the overall model complexity since the model does not need to predict actual table names for the FROM-clause.

Furthermore, as explained before in Section \ref{sec:runtime}, for single-table queries our translation model sometimes produces erroneous SQL queries where the table name in the FROM-clause does not match the attribute names used by the query. 
These errors are handled in the post-processing step and require to infer the correct table names from the attributes used in the SQL query.
Thus increasing the model complexity to predict also the join paths and table names even increases the rate of errors that would need to be handled in the post-processing phase.

Hence, as explained before instead of predicting all tables and the join paths using the model, the model only outputs the \textit{@JOIN} placeholder.
\system{}'s post-processing phase then uses the schema information in order to infer the table names and a join path from the attributes in the SQL output of the model.
In case multiple join paths are possible to connect all the required tables, we select the join path that is minimal in its length.
In the future, we want to provide a better way to infer the join path from semantics of the input NL query using a separate model that is only used to expand the \textit{@JOIN} placeholder.
Furthermore, we could combine the current schema-based and  model-based approaches for expanding the \textit{@JOIN} placeholder to handle cases where one or the other approach fails.

\subsection{Nested Queries}

Handling arbitrary nested queries in an NLIDB is a hard task on its own. 
In the current version of \system{}, we only handle a limited subset of possible SQL nestings by extending training data as follows.
The main idea is to add additional templates that represent common forms of nested queries where the slots for the outer and inner query can be instantiated individually.
An example for a NL-SQL template pair looks as follows:\\

\makebox[0.45\textwidth][c]{
\centering
\begin{minipage}[c]{0.45\textwidth}
\noindent \textbf{SQL Template:} \\
\textit{Select \textsf{\{Attribute\}(s)} From \textsf{\{Table\}} Where
(Select \{MaxMinAttribute\} From \textsf{\{Table\} Where \textsf{\{Filter\}})})}
\vspace{1.5ex}\\
\textbf{NL Template:} \\
\textit{\{SelectPhrase\} the \textsf{\{Attribute\}(s)}  
\{FromPhrase\} \textsf{\{Table\}} \{WherePhrase\} \textsf{\{MaxMinAttribute\}}}\\
\end{minipage}}

This template is then instantiated during the first phase of the data generation process.
For example, the following pair of instantiated queries could be generated for the training set from the template pair before:\\

\makebox[0.45\textwidth][c]{
\centering
\begin{minipage}[c]{0.45\textwidth}
\noindent \textbf{SQL Query (Training Set):} \\
\emph{SELECT name FROM mountain where height=  
(SELECT MAX(height) FROM mountain WHERE state=@STATE.NAME)}
\vspace{1.5ex}\\
\textbf{NL Query (Training Set):} \\
\emph{What is name of the mountain with maximal height in @STATE.NAME}\\
\end{minipage}}

The instantiated queries are augmented automatically in the same way as we do it for non-nested queries.
In its current version, \system{} only supports uncorrelated nestings in the WHERE-clause using different types keywords such as EXISTS and IN, as well as nested queries where the inner query returns an aggregate results as shown before.
The capabilities of \system{} could be extended by adding further templates that are instantiated in the first phase of the data generation.

%% file: 06_evaluation.tex
\section{Experimental Evaluation}
\label{sec:evaluation}

The main goal of the experimental evaluation is to compare the capabilities of \system{} to translate NL into SQL with those of other Natural Language Interfaces for Databases (NLIDBs).
In this section, we analyze the performance of \system{} compared to two other baselines: NaLIR \cite{nalir_sigmod14} and a recent approach in \cite{DBLP:conf/acl/IyerKCKZ17} that also uses a Recurrent Neural Network model. We perform evaluations on two benchmarks datasets before reporting the results of our user study.

In order to compare \system{} to other baselines we use the GeoQuery benchmark that has already been used to evaluate other NLIDBs.
However, this benchmark does not explicitly test different linguistic variations which is important to understand the robustness of an NLIDB.
For testing different linguistic variants in a principled manner, we therefore curated a new benchmark as part of this paper that covers different linguistic variations for the user NL input and maps it to an expected SQL output. The benchmark is available online \footnote{\url{https://datamanagementlab.github.io/ParaphraseBench/}}. 

In the following, we first present the design of a new benchmark and then discuss the details of other benchmarks we used as well as the details of the setup for the other systems we used as baselines.
Afterwards, we present the results of running these benchmarks.
In the last experiment, we also show the efficiency of our optimization procedure that automatically tunes the parameters to instantiate the training data for \system{} to find an optimal configuration.

\vspace{3ex}
\subsection{A New Benchmark}

The schema of our new benchmark models a medical database which contains only one table comprises of hospital's patients attributes such as name, age, and disease.
In total, the benchmark consists of $290$ pairs of NL-SQL queries.
The queries are grouped into one of the following categories depending on the linguistic variation that is used in the NL query: 
\naive{}, syntactic paraphrases, morphological paraphrases, and lexical paraphrases as well as a set of queries with missing information.

While the NL queries in the \naive{} category represent a direct translation of their SQL counterpart, the other categories are more challenging:  syntactic paraphrases emphasize structural variances, lexical paraphrases pose challenges such as alternative phrases, semantic paraphrases use semantic similarities such as synonyms, morphological paraphrases add affixes, apply stemming, etc., and the NL queries with missing information stress implicit and incomplete NL queries.

In the following, we show an example query for each of these categories in our benchmark:

\begin{itemize*}
\item \textbf{\Naive{}}: \emph{``What is the average length of stay of patients where age is 80?"}
\item \textbf{Syntactic:} \emph{``Where age is 80, what is the average length of stay of patients?"}
\item \textbf{Morphological:} \emph{``What is the averaged length of stay of patients where age equaled 80?"}
\item \textbf{Lexical:} \emph{``What is the mean length of stay of patients where age is 80 years?"}
\item \textbf{Semantic:} \emph{``What is the average length of stay of patients older than 80?"}
\item \textbf{Missing Information:} \emph{``What is the average stay of patients who are 80?"}
\end{itemize*}

\subsection{Other Benchmarks}

In addition to the `Patients' benchmark dataset mentioned before, we also measure the performance of our method on a database of US geography along with 280 test examples which first introduced by Mooney \cite{Zelle1996LearningTP}. We used Geo880-version (called `Geo' in this paper) introduced by \cite{DBLP:conf/acl/IyerKCKZ17} which comprises of pairs of natural language and SQL queries, instead of the original version which correspond natural language with its logical form. Geo consists of 7 tables which represent geographical entities such as US states, cities, rivers and mountains. 
In addition to queries over individual tables, this benchmark  includes more complex queries such as joins, nested queries, and questions that are in general semantically harder than single-table queries with and without aggregation.

For the sake of comparison, our approach was evaluated using the same 280-pair testing set as \cite{DBLP:conf/acl/IyerKCKZ17}, who used the other 600 for training and validation. To better understand how our model behaves across different types of queries, we categorized a large amount of the testing set into various SQL forms and patterns of queries:

\begin{itemize*}
    \item 44 queries using the SQL \emph{IN} operator, generally for comparison of values across two or more tables.
    \item 11 queries requiring aliases and the SQL \emph{as} operator.
    \item 91 `argmin' or `argmax' queries such as `get the state with the highest population.' The SQL form generally follows the pattern \emph{SELECT \dots FROM \{TABLE1\} WHERE \{COL\} = (SELECT max(\{COL\}) FROM \{TABLE2\})}.
    \item 11 two-table join requests of the form \emph{SELECT \dots from \{TABLE1\},\{TABLE2\} WHERE \dots}
    \item 29 queries containing the SQL `GROUP BY' operator.
\end{itemize*}

\subsection{Setup and Baselines}

Based on the benchmarks mentioned before, we have evaluated our system against recent NLIDB approaches:

As first baseline we compared against the neural semantic parser introduced by \cite{DBLP:conf/acl/IyerKCKZ17}, which also uses a deep neural network for translating NL to SQL. 
However, \cite{DBLP:conf/acl/IyerKCKZ17} inherently depends on manually annotated training data on each schema in addition to on-the-fly user feedback.
In order to bootstrap the training data creation, \cite{DBLP:conf/acl/IyerKCKZ17} provide a limited set of templates to generate some initial NL-SQL pairs.
This corresponds to the template-only approach as presented in \cite{DBLP:conf/acl/IyerKCKZ17}. We first compared the performance of our approach to the accuracy of their model trained on only the context-independent training set generated from their templates. We name this version of their model \emph{NSP} for `Neural Semantic Parser.'
We also evaluate \cite{DBLP:conf/acl/IyerKCKZ17}'s model when using the extended training set that additionally contains manual curated NL-SQL pairs; we name this version \emph{NSP++}.

As a second baseline, we used NaLIR \cite{nalir_sigmod14}, which upon its introduction was deemed as the state-of-the-art natural language interface for databases. On the core of its database-agnostic framework is a rule-based system which adjusts and refines a dependency parse tree to a query-tree, based on the tree structure and the mapping between its nodes to the RDBMS elements.
NaLIR relies heavily on an user feedback to refine word to SQL element/function mapping which later determines the final SQL string output.
To allow fair comparison between our system and the other baseline, we run the NaLIR implementation in a non-interactive setting.

\subsection{Exp. 1: Overall Results}

We evaluated the performance of all NLIDB systems in terms of their accuracy, defined as the number of natural language queries translated correctly over the total number of queries in the test set. The correctness of a resulting SQL query is determined by whether the yielded records from its execution to the RDBMS contain the information that is asked by the query intent. The correctness criteria is relaxed by considering the execution result which consist of columns supersets of the queried column to be correct. We argue that in practice, users are still able retrieve the desired information by examining the entire columns of returned rows. Imagine a scenario depicted by Figure \ref{fig:screenshot} where a user intends to retrieve the age of patients who diagnosed with flu. By looking at the age column, the user is able to yield a satisfying answer.

Table \ref{table:results} summarizes the accuracy measures of all NLIDB systems on our two benchmark datasets using already previously defined correctness criteria.
We can see that \system{} outperforms both systems which require no manual effort to support a new database (NaLIR and NSP).
NSP and NaLIR both have a low accuracy on both data sets.
While NaLIR's rule-based approach fails in many cases to produce a direct translation without involving a user,  NSP can not translate most of the queries correctly due to its limited training set size.
Furthermore, we see that NSP++ which requires a costly supervised creation of a training set of NL-SQL pairs achieves the highest accuracy.
It is compelling that \system{}, which does not require manual effort, similar to NSP and NaLIR, is able to come closest to the accuracy of NSP++. 
Moreover, it is interesting to note that NSP++ achieves a high quality on Geo while \system{} only achieves approximately 50\%.
However, after analyzing the training data of NSP++ we saw that it contains many of the structurally similar NL-SQL pairs in the test set and thus the learned model is heavily over-fitted. 
A last observation is that \system{} performs worse on Geo than on Patients. 
The reason is that Geo contains a huge fraction of rather complex join and nested queries that are currently not supported in \system{}.

\begin{table*}[]
\centering
\begin{tabular}{|l|l|l|}
\hline
& \textbf{Patients} & \textbf{Geo} \\ \hline
\begin{tabular}[c]{@{}l@{}}\textbf{NaLIR} \textbf{(w/o feedback)}\end{tabular} & 5.49\% & 7.14\%\\ \hline
\textbf{NSP++} &N/A &83.9\%        \\ \hline
\begin{tabular}[c]{@{}l@{}}\textbf{NSP} \textbf{(template only)}\end{tabular} & 10.6\% &5.0\%       \\ \hline
\textbf{\system{}} & 75.93\% & 48.6\%          \\ \hline
\end{tabular}
\caption{Accuracy comparison between \system{} and other baselines on the two benchmark datasets}
\label{table:results}
\end{table*}

\subsection{Exp. 2: Performance Breakdown}

In this experiment, we show the result breakdown of using the Patient benchmark in Table \ref{table:breakdown}. 
We see that \system{} outperforms the other two approaches for each test category in our benchmark.

The first baseline (NSP), which also uses a neural machine translation that is trained on data instantiated by only a very limited number of templates, achieves less than $10\%$ accuracy on average for all test categories only.
We can also see that the accuracy for the more challenging test sets (Lexical, Semantic, Missing Information) 
is lower than the one for the \naive{} test set.

NaLIR \cite{nalir_sigmod14} interestingly also only achieves a slightly higher accuracy of $10/57=17.5\%$ for the \naive{} testing set, and even lower on other paraphrased test sets.
Unlike ours, NaLIR relies on off-the-shelf dependency parser library which performs reasonably good on well-structured sentences.
Most of NaLIR's failure cases are due to dependency parsing errors caused by ill-formed, incomplete, or keywords-like queries.
Moreover, NaLIR often fails to find correct candidate mapping of query tokens to RDBMS elements due to the paraphrased input.

\begin{table*}[]
\centering
\small
\begin{tabular}{|l|l|l|l|l|l|l|l|}
\hline
 & \textbf{Naive} & \textbf{Syntactic} & \textbf{Lexical} & \textbf{Morphological} & \textbf{Semantic} & \textbf{Missing} & \textbf{Mixed}\\ \hline
\begin{tabular}[c]{@{}l@{}}\textbf{NaLIR} \textbf{(w/o feedback)}\end{tabular} & 7.01\% & 10.5\% & 5.2\% & 7.01\% & 0\% & 1.75\% & 7.01\%\\ \hline
\begin{tabular}[c]{@{}l@{}}\textbf{NSP} \textbf{(template only)}\end{tabular} & 19.29\% & 7.01\% & 5.2\% & 17.54\% & 12.96\% & 3.5\% & 8.7\% \\ \hline
\textbf{\system{}} &96.49\% & 94.7\% &  75.43\% & 85.96\% & 57.89\% & 36.84\% & 84.2\%\\ \hline
\end{tabular}
\caption{Accuracy break-down between \system{} and other baselines for the Patient benchmark}
\label{table:breakdown}
\end{table*}

\subsection{Exp. 3: Efficiency of Optimization}

As described in Section 3.3, we applied Bayesian Optimization to the hyperparameters of our model's training set generator. This technique is useful for global optimization problems that evaluate expensive black-box functions; as such it has become popular for optimizing deep learning models that take in a seemingly arbitrary setup of hyperparameters such as the number of layers or perceptrons per layer of a convolutional neural network (CNN). However, instead of applying it to our seq2seq model, we extrapolate one step backwards and attempt to optimize the nature of the training set to which the model will be exposed.
In \system{}, we use the optimization procedure provided by \cite{gpyopt2016} to run the process. This allowed for implementation of both continuous and discrete variables. 

In this experiment, we show the results of applying the optimization procedure for generating the training data for the Geo-benchmark. 
As a test workload, we used the GeoQuery query testing set of 280 pairs provided by \cite{DBLP:conf/acl/IyerKCKZ17}.
We defined bounds upon each variable based upon their function; initial parameter samples were drawn from the feasible region created in this way.
The procedure works as follows: 
it used the `Expected Improvement' acquisition function first introduced by \cite{DBLP:conf/ifip7/Mockus74} to sample new configurations in every round (called batch). 
These configurations where then trained for $10$ epochs.
At the end, after all batches finished, the configuration which achieved the highest accuracy is returned. 

\begin{figure}
\centering
\small
\includegraphics[width=0.4\textwidth]{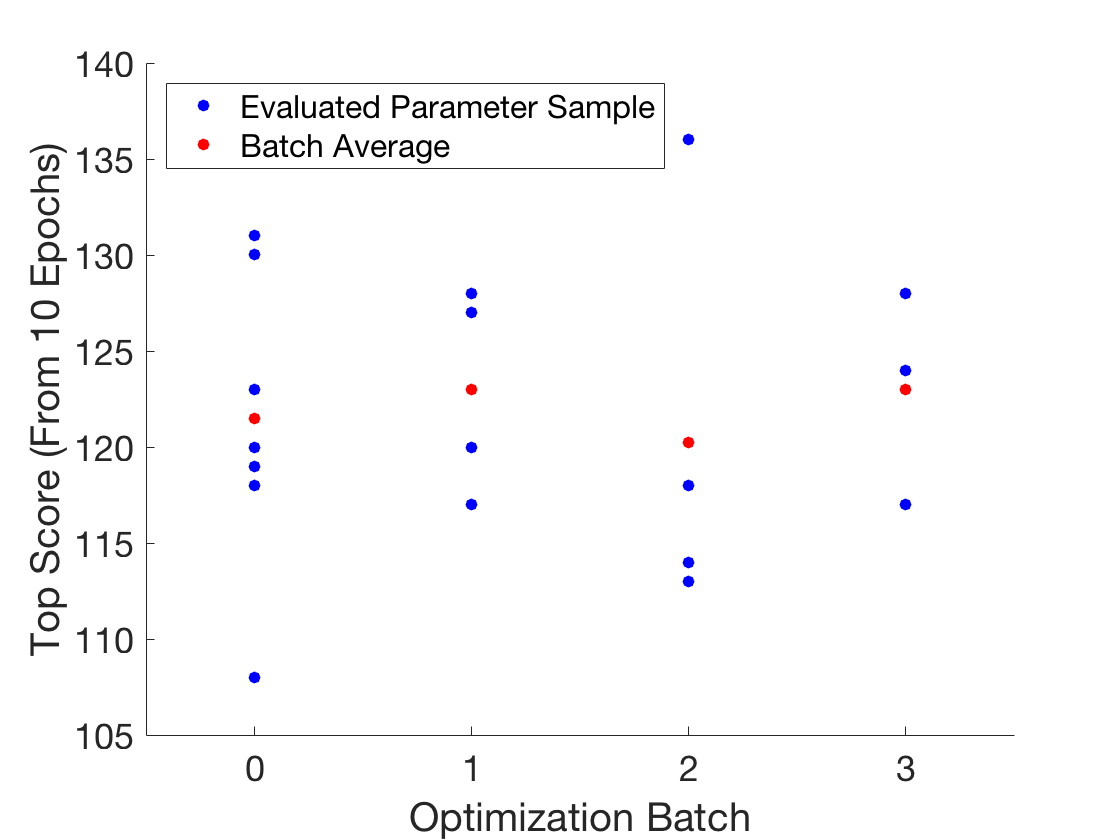}
\vspace{2ex}
\caption{Accuracy Results of Bayesian Optimization}
\label{fig:3_3}
\end{figure}

Figure \ref{fig:3_3} shows the accuracy recorded over $4$ batches with $4$ sampled configurations per batch. 
For the first round (batch $0$), we used $8$ configurations (sampled randomly from the feasible space) to have a higher exploitation and enable a richer choice for the Bayesian acquisition function.
The highest recorded configuration over all batches had an accuracy of $136/280$ and thus successfully translated $29$ queries more than the model with the lowest accuracy.
The best configuration we found by the optimization procedure was the one we used to generate the training data for the model in experiment 2 for the Geo benchmark.

%% file: 07_userstudy.tex
\section{User Study}
\label{sec:study}

In addition to the evaluation using benchmarks, we also conducted a user study in which participants were asked to complete a set of tasks that could help us evaluate the ability of NLIDB systems in translating arbitrary NL queries into SQL. 
In the user study, we only compared \system{} and NaLIR since both can be executed on arbitrary schemata without any manual effort of curating a training set.
We excluded NSP since it can only provide low accuracy as shown in the evaluation before when using the template-based approach and NSP++ because it requires high manual efforts to curate a manual training set.
For a fair comparison, we disabled the interaction modules in \system{} (i.e., auto-completion) and NaLIR (i.e., user feedback) to compare only the direct translation capabilities. 
The details of the experiment are covered in the following two sections.

\subsection{Experimental Setup}

In the following, we describe the setup of the user study.

   \paragraph*{Users} 13 people participated in our experiments that were recruited with flyers posted on our university campus. The users had a variety of backgrounds but we required that all of them had sufficient understanding of SQL and were able to distinguish between a correct and a wrong translation.
    
    \paragraph*{Datasets}
    We used the two datasets that we also used before in the evaluation: 1 -- Patients dataset, and 2 -- the Geo dataset. 
    
    \paragraph*{Scenario}
    The user study consisted of three steps. In the first step, the participants were allowed to explore the tables shown on the system interface for both the \textit{Patients} and \textit{Geoquery} databases. After getting familiar with the data schema including table and column names, we asked participants to come up with 20 questions (10 for each schema) that they wanted to ask from the system and record these questions offline. The goal of proposing questions beforehand (as opposed to at the query time) was to prevent sequential or experiential bias that might have occurred. Such bias might be entailed as a result of observing more precise translations for particular types of questions (e.g, when the systems can better handle \textit{select} queries of certain type). The participant might then propose similar question types and fail to span a sufficiently wide range of questions. 
    In the second step, the participants chose one of the databases to begin the evaluation and sequentially entered the questions that they had listed in step one. Since in this study we are interested in comparing two systems, users received two responses (returned by DBPal and NaLIR) each time they entered a question. In order to prevent bias, we anonymized the systems names for the users (in the form of system \textit{A} vs. system \textit{B}). The responses from systems included a generated SQL translation alongside a tabular view of the returned result. After entering each question and observing the results, users were asked to grade the quality of the translations in a scale of 1-5. The following explains how they graded after receiving the results: 
    \begin{itemize*}
    \item \textbf{Scenario 1:} 
    Both systems return a correct response in the first attempt: in this case, both of them receive 5 points. 
    \item \textbf{Scenario 2:} 
    At least one of the systems fails: in this case, the user can rephrase their question for at most 4 more times or until both systems produce a correct translation. Once both of the systems produce correct translations, or the maximum number of rephrase attempts are exceeded, the user enters a grade between 1 and 5 depending on the fraction of correct translations; i.e., for every wrong translation the user subtracts a point. For example, if the user gets the correct answer after 3 re-attempts (i.e., 2 failed re-attempts)  the user will give in total $5-2 = 3$ points. 
    \end{itemize*}
    
    Afterwards in the third and final step, the user repeats the evaluation procedure on the second dataset/schema.
    
    It is worth mentioning that the users had no prior knowledge about the schema. Furthermore, we did not provide them with any sample NL questions (including SQL samples) since they could have caused bias.
    
    \paragraph*{Collected data} In order to perform a richer post-experimental analysis on the data, we recorded not only the grades for each question but also fine-grained details such as the actual questions, their rephrase/reword attempts and the results returned by the systems. This data allows us to evaluate the performance of two systems based on the types of questions and the level of difficulty in each category. 
 
 \subsection{Result Summary}
 \label{US:results}
 
\begin{figure*}
\hspace{12ex}
\makebox[\textwidth][c]{
\begin{minipage}[c]{\textwidth}
\subfloat[Patients Dataset\label{fig:grades1}]
{
    \begin{tikzpicture}[scale=0.7]
        \begin{axis}[
        ybar,
        enlargelimits=0.15,
        symbolic x coords={1, 2, 3, 4, 5},
        legend style={at={(0.5,-0.15)},
        anchor=north,legend columns=-1},
        nodes near coords,
        nodes near coords align={vertical},
        cycle list = {black,black!70,black!40,black!10}
        ]
        \addplot+[] coordinates {(1, 35) (2, 20) (3, 20) (4, 13) (5, 42)};
        \addplot+[fill,text=black] coordinates {(1, 104) (2, 9) (3, 6) (4, 4) (5, 7)};
        \legend{\system{}, NaLIR}
        \end{axis}
    \end{tikzpicture}
}
\subfloat[Geo Dataset\label{fig:grades2}]
{
        \begin{tikzpicture}[scale=0.7]
        \begin{axis}[
        ybar,
        enlargelimits=0.15,
        symbolic x coords={1, 2, 3, 4, 5},
        legend style={at={(0.5,-0.15)},
        anchor=north,legend columns=-1},
        nodes near coords,
        nodes near coords align={vertical},
        cycle list = {black,black!70,black!40,black!10}
        ]
        \addplot+[] coordinates {(1, 40) (2, 17) (3, 25) (4, 12) (5, 36)};
        \addplot+[fill,text=black] coordinates {(1,  116) (2, 3) (3, 2) (4, 1) (5, 8)};
        \legend{\system{}, NaLIR}
        \end{axis}
        \end{tikzpicture}
}
\end{minipage}}
\caption{User ratings for different data sets on \system{} and NaLIR (higher is better)}
\label{fig:grades}
\end{figure*}
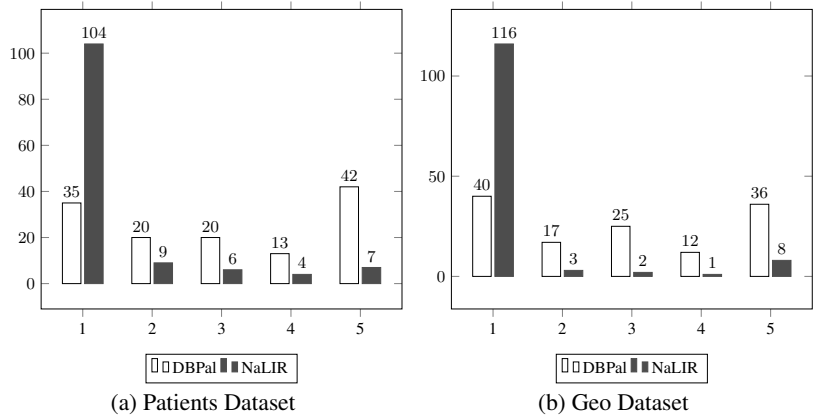

 Figure \ref{fig:grades1} and Figure \ref{fig:grades2} show the grades received by the two NLIDB systems for both the \textit{Patients} and \textit{Geo} database. In this figure, the $x$ axis represents the grade and the $y$ axis represents the number of times the given translator has received that grade from the users.
 The average grades on both databases for DBPal and NaLIR are $2.98$ and $1.4$, respectively (i.e., the higher the better).
 
The distribution of the questions asked by the participants can be summarized according to the following categories: 1 -- single-table \textit{select} statements excluding the ones with aggregation or \textit{group by}  ($24\%$), 2 -- complex questions requiring join and nest operations ($26\%$), 3 -- \textit{select} statements with aggregation ($39\%$), 4 -- \textit{select} statements with \textit{group by} ($6\%$), and 5 -- the others including ambiguous, top $k$, binary yes-no questions, and other rare types of questions ($5\%$). Table \ref{table:user_study} summarizes the percentage of each category of questions that could be correctly translated by \system{} and NaLIR within at most 5 rephrase attempts.

\begin{table*}[h!]
\centering
\small
\begin{tabular}{|c c c c c c|} 
 \hline
 \textbf{Translator} & \textbf{Category 1} & \textbf{Category 2} & \textbf{Category 3} & \textbf{Category 4} & \textbf{Category 5}\\ [0.5ex] 
 \hline
 \system{} & 70\% & 35\% & 87\% & 83\% & 0\%\\  \hline
 NaLIR & 38\% & 0\% & 33\% & 17\% & 0\%\\ [1ex] 
 \hline
\end{tabular}
\caption{Percentage of correct translations by \system{} vs. NaLIR for each category of questions explained in Section \ref{US:results}.}
\vspace{-2.5ex}
\label{table:user_study}
\end{table*}

According to the results illustrated in Table \ref{table:user_study}, NaLIR suffers more in handling complex questions requiring join and nest operations. 
Furthermore, we also saw that NaLIR is usually unable to translate an NL query correctly when the request is entered as a set of keywords as opposed to a complete question.

%% file: 08_related.tex
\section{Related Work}
\label{sec:related}

Previous work on Natural Language Processing (NLP) has heavily relied on classical statistical models to implement tasks such as semantic parsing that aim to map a natural language utterance to an unambiguous and executable logical form \cite{Zelle1996LearningTP}. 
More recent results on semantic parsing such as \cite{Jia2016DataRF, Dong2016LanguageTL} have started to employ deep recurrent neural networks (RNNs), particularly sequence-to-sequence architecture, to replace traditional statistical models. 
RNNs have shown promising results and outperform the classical approaches for semantic parsing since they make only few domain-specific assumptions and thus require only minimal feature engineering.

An important research area to support non-experts to specify ad-hoc queries over relational databases are keyword search interfaces are widely used by \cite{DBLP:journals/debu/YuQC10}. 
Recently, there have been extensions to keyword-based search interfaces to interpret the query intent behind the keywords in the view of more complex query semantics \cite{DBLP:conf/sigmod/TataL08,DBLP:journals/pvldb/BlunschiJKMS12,DBLP:journals/pvldb/BergamaschiGILV13}. 
In particular, some of them support aggregation functions, boolean predicates, etc. 

Another relevant area for this paper are Natural Language Interfaces for Databases (NLIDBs).
NLIDBs have been studied in the database research community since the 1990's \cite{precise2,nalir_sigmod14,athena_vldb16,Androutsopoulos1995NaturalLI}. 
Most of this work relied on classical techniques for semantic parsing and used rule-based approaches for the translation into SQL.
However, these approaches have commonly shown poor flexibility for the users who phrase questions with different linguistic styles using paraphrases and thus failed to support complex scenarios.

More recent approaches tackled some of the limitations of the original NLIDBs. 
For example, the system ATHENA \cite{athena_vldb16} relies on a manually crafted ontology that is used to make query translation more robust by taking different ways of phrasing a query into account. Yet, since ontologies are domain-specific, they need to be hand-crafted for every new database schema.  On the other hand, the system NaLIR \cite{nalirvldb} relies on a off-the-shelf dependency parser that could also be built on top of a deep model. However, it still implements a rule-based system that struggles with variations in vocabulary and syntax. Our system attempts to solve both of those issues by being domain independent as well as robust to grammatical alterations. 

Furthermore, some recent approaches leverage deep models for end-to-end translation similar to our system (e.g., \cite{DBLP:conf/acl/IyerKCKZ17}).
However, a main difference of our system to \cite{DBLP:conf/acl/IyerKCKZ17} is that their approach requires manually handcrafting a training set for each novel schema/domain that consist of pairs of natural language and SQL queries. 
In contrast, our approach does not require a hand-crafted training set.
Instead, inspired by \cite{Wang2015BuildingAS}, our system generates a synthetic training set that requires only minimal annotations to the database schema. 

Another recent paper that also uses a deep model to translate NL to SQL is \cite{DBLP:journals/corr/abs-1711-04436}.
First, the approach in this paper is a more classical approach based on identifying the query intend and then filling particular slots of a query.
In their current version, \cite{DBLP:journals/corr/abs-1711-04436} can only handle a much more limited set of NL queries compared to \system{}.
Furthermore, the approach in \cite{DBLP:journals/corr/abs-1711-04436} leverages reinforcement learning to learn from user feedback in case the query could not be translated correctly, which is an orthogonal issue that could also be applied to \system{}.

Finally, none of the above-mentioned approaches combine their translation pipelines with additional functions such as auto-completion.
These features not only make query formulation easier by helping users to phrase questions even without knowing the database schema, but they also help users to write less ambiguous queries that can be more directly translated into SQL.

%% file: 09_conclusion.tex
\section{Conclusions \& Future Work}
\label{sec:concl}

The current prototype of \system{} already shows a significant improvement over other state-of-the-art-systems such as NaLIR \cite{nalirvldb} when dealing with paraphrasing and other linguistic variations. 
Furthermore, compared to other recent NLIDB approaches that leverage deep models for the query translation from NL to SQL, \system{} only requires minimal manual effort.
At the moment the main limitation of \system{}  is the lack of coverage to explain results to the user and ways to correct the queries if the translation was inaccurate. 

A future avenue of development is therefore to allow users to incrementally build queries in a chatbot-like interface, where the system can ask for clarifications if the model cannot translate a given input query directly. We expect that this feature will also be especially helpful for building complex nested queries in an incremental manner. 
Furthermore, integration with other deep models (e.g., for Ques\-tion-Answering) seems to be another promising avenue for future exploration to being able to handle more query types that are less SQL-like.
Finally, we also plan to further extend the training data instantiation and augmentation phase with additional templates and lexicons as well as additional support for even more complex linguistic variations.